\documentclass{optica-article}

\journal{opticajournal} 

\articletype{Research Article}

\usepackage{lineno}

\usepackage{hyperref}
\usepackage{float}
\usepackage{graphicx}
\usepackage{subcaption}
\usepackage{caption}
\usepackage{physics}
\usepackage{microtype}
\usepackage[inter-unit-product={\;},output-decimal-marker={.},per-mode=symbol-or-fraction,sticky-per,exponent-product = \cdot, output-product = \cdot,separate-uncertainty=true]{siunitx}
\usepackage[wby]{callouts}
\usepackage{amsfonts}
\usepackage{amsmath}
\usepackage{tabularx}
\usepackage[title,toc,titletoc,page]{appendix}

\begin{document}

\title{Time-bin entanglement at telecom wavelengths from a hybrid photonic integrated circuit}

\author{Hannah Thiel,\authormark{1,*} Lennart Jehle,\authormark{2,3} Robert J. Chapman,\authormark{1,4} Stefan Frick,\authormark{1} Hauke Conradi,\authormark{3} Moritz Kleinert,\authormark{3} Holger Suchomel,\authormark{5} Martin Kamp,\authormark{5} Sven Höfling,\authormark{5} Christian Schneider,\authormark{5,6} Norbert Keil,\authormark{3} and Gregor Weihs\authormark{1}}

\address{\authormark{1}Institut f\"{u}r Experimentalphysik, Universit\"{a}t Innsbruck, 6020 Innsbruck, Austria\\
\authormark{2}Faculty of Physics \& Vienna Doctoral School in Physics \& Vienna Center for Quantum Science and Technology, University of Vienna, 1090 Vienna, Austria\\
\authormark{3}Fraunhofer Institute for Telecommunications, Heinrich-Hertz-Institut, 10587 Berlin, Germany\\
\authormark{4}Optical Nanomaterial Group, Institute for Quantum Electronics, Department of Physics, ETH Zurich, 8093 Zurich, Switzerland\\
\authormark{5}Technische Physik, Universit\"{a}t W\"{u}rzburg, 97074 W\"{u}rzburg, Germany\\
\authormark{6}Institute of Physics, University of Oldenburg, 26129 Oldenburg, Germany\\}

\email{\authormark{*}hannah.thiel@uibk.ac.at} 


\begin{abstract*}
Mass-deployable implementations for quantum communication require compact, reliable, and low-cost hardware solutions for photon generation, control and analysis.
We present a fiber-pigtailed hybrid photonic circuit comprising nonlinear waveguides for photon-pair generation and a polymer interposer reaching $\SI{68}{\deci\bel}$ of pump suppression and photon separation with  $>\SI{25}{\deci\bel}$ polarization extinction ratio.
The optical stability of the hybrid assembly enhances the quality of the entanglement, and the efficient background suppression and photon routing further reduce accidental coincidences.
We thus achieve a $\left(96_{-8}^{+3}\right)\%$ concurrence and a $\left(96_{-5}^{+2}\right)\%$ fidelity to a Bell state.
The generated telecom-wavelength, time-bin entangled photon pairs are ideally suited for distributing Bell pairs over fiber networks with low dispersion.
\end{abstract*}

\section{Introduction}
As data traffic continues to grow, the cryptography community is increasingly aware of the importance of methods and devices ensuring an efficient and secure data transmission.
For the required security, the conventional public key infrastructure has been shown to be unsuitable in the long term~\cite{Bennett2000}.
Quantum communication, in contrast, provides information-theoretical security when implemented correctly~\cite{Renner2005}.
A multitude of implementations have been demonstrated in field trials using metro networks~\cite{Chen2021,Dynes2019,Chen20212}. 
Among those, the majority do not rely on entanglement and the experimental setups have the size of a computer rack or larger.
For mass-deployment and practical implementation, however, quantum communication systems must become more compact, cost-effective and scalable.
This can be achieved via quantum system-on-chip modules~\cite{Wang2020,Sibson2017,Caspani2017}.
Also allowing for individual optimization of dissimilar components, hybrid photonic integrated circuits (PICs) have recently received much attention in quantum photonics~\cite{Elshaari:20,Kim:20}, where challenges ranging from single-photon generation to reconfigurable photon routing and high-efficiency detection have particularly demanding requirements that cannot be fulfilled by a single photonic platform.

In addition to scaling up quantum communication systems, one must strive for more than conditional security.
It will be difficult to certify all quantum communication source and receiver modules and to ensure their long-term integrity.
Therefore, entanglement-based quantum key distribution (QKD) schemes are  promising, especially when used in future device-independent schemes that rely on the principle of non-locality and can generate secure keys even for untrusted devices~\cite{Ekert1991,Barrett2005,Pironio2009,Zapatero2023}.

For pratical QKD, the transmitted qubits must be compatible with the existing telecom infracture and also conserve the entanglement en route.
To this end, time-bin entanglement is especially well suited as it does not suffer from decoherence due to polarization mode dispersion~\cite{Thew2002,Antonelli2011}.
In this scheme, a photon pair is created in a coherent superposition of two time bins and the communicating parties each receive one of the photons allowing them to test the quality of the entanglement and generate bits of a shared secret key.
A number of experiments have demonstrated this form of entanglement as a proof-of-principle for entanglement-based QKD~\cite{Marcikic2004}, using integration-ready sources~\cite{Tanzilli2002,Jayakumar2014,Chen2018}, generating on-demand time-bin qubits~\cite{Ilves2020}, or achieving a distance record~\cite{Yin2016}. 
However, few telecom time-bin entanglement sources have been realized on-chip or in optical fiber~\cite{Takesue2014, Xiong:15, Zhang2018}.

We present in this article an on-chip, partially fiber-pigtailed source of time-bin entangled photon pairs in the telecom wavelength range working at room temperature.
The photon pairs are generated in a nonlinear crystal made of aluminum gallium arsenide, called a Bragg-reflection waveguide (BRW)~\cite{Yeh1976,Helmy2006,Appas2022}.
This source is integrated with a polymer chip, the PolyBoard, which hosts all passive optical components including a long-pass filter (LP) showing $\SI{68}{\deci\bel}$ of pump suppression, a polarizing beam splitter achieving $>\,$\SI{25}{\deci\bel} polarization extinction ratio $\left(P\!E\!R\right)$, and specially designed grooves for fiber pigtailing~\cite{Kleinert:15,Kleinert:17,Zhang:16}.
We achieve a coincidence rate of 460\,Hz per mW continuous-wave (CW) external pump power between the signal and idler photons without correcting for fiber loss or detector efficiency.
In the time-bin entanglement scheme, this results in photon pair rates of \SI{1.4}{\hertz} per mW of external pump power, a concurrence of $\left(96_{-8}^{+3}\right)\%$ and a fidelity of $\left(96_{-5}^{+2}\right)\%$ to the $\ket{\Phi^+}$ Bell state.

The article is structured as follows:
After a brief explanation of the time-bin scheme, both the BRW and the PolyBoard are introduced in more detail followed by a section on their hybrid integration and assembly process.
We then perform a classical characterisation of the PIC and finally present the time-bin measurements including state tomography using maximum likelihood estimation~\cite{James2001,Takesue2009}.

\section{Materials and methods}
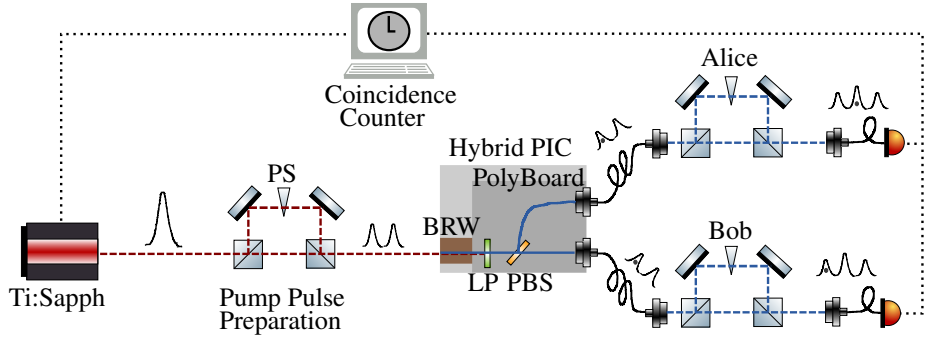
\begin{figure}[!htbp]
	\captionsetup{singlelinecheck = false, justification=raggedright}
    	 \centering
    	 \begin{annotate}{\includegraphics[width=0.9\textwidth]{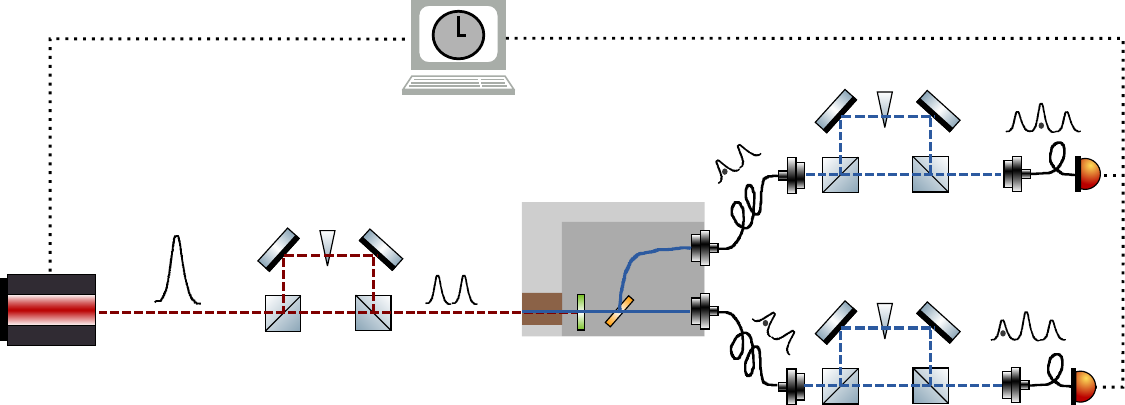}}{1}
    	\draw[very thick,black] (-6.3,-1.5) node[anchor=north west]{Ti:Sapph};
        \draw[very thick,black] (-0.8,-0.55) node[anchor=north west]{BRW};
        \draw[very thick,black] (-0.13,0.1) node[anchor=north west]{PolyBoard};
        \draw[very thick,black] (-0.45,0.5) node[anchor=north west]{Hybrid PIC};
        \draw[very thick,black] (-0.2,-1.3) node[anchor=north west]{LP};
        \draw[very thick,black] (0.32,-1.3) node[anchor=north west]{PBS};
        \draw[very thick,black] (-2.1,1.2) node[anchor=north west]{Coincidence};
        \draw[very thick,black] (-1.9,0.9) node[anchor=north west]{Counter};
        \draw[very thick,black] (-3.5,-1.5) node[anchor=north west]{Pump Pulse};
        \draw[very thick,black] (-3.5,-1.8) node[anchor=north west]{Preparation};
        \draw[very thick,black] (2.9,1.7) node[anchor=north west]{Alice};
        \draw[very thick,black] (3,-0.6) node[anchor=north west]{Bob};
        \draw[very thick,black] (-2.86,0.15) node[anchor=north west]{PS};
    	 \end{annotate}
        \caption{\textbf{Time-bin entanglement scheme.} Our setup includes the preparation of pump pulse pairs, the PIC, where the telecom photon pairs are created, filtered, separated and coupled into fiber, as well as the stations of Alice and Bob, where the entanglement is analyzed. These consist of interferometers with variable phase shift (PS) and single photon detectors.}
        \label{fig:TimeBin_Schematic}
\end{figure}

We implement the time-bin entanglement as illustrated in Fig.~\ref{fig:TimeBin_Schematic}.
A coherent superposition of time bins is prepared by passing a pulsed Ti:Sapphire laser with \SI{76}{\mega\hertz} repetition rate and \SI{0.8}{\nano\meter} bandwidth emitting at \SI{767}{\nano\meter} through an asymmetric free-space Michelson interferometer.
This splits each pulse into an early and a late time bin separated by a \SI{3}{\nano\second} delay and the pulse pair then travels to the hybrid PIC.
One photon pair is produced with probability $p\ll1$ by either the early or late pump pulse, separated by polarization and routed to two optical fibers on the hybrid PIC.
The photons are sent to two parties, Alice and Bob, who analyze the entanglement via interferometers with the same delay as the pump interferometer.
In our setup, all three interferometers are folded into the same physical interferometer.
Finally, the photons are measured by superconducting-nanowire-single-photon detectors (SNSPDs) with \SI{40}{\pico\second} timing jitter and $>\,$\SI{60}{\percent} detection efficiency.
A triple coincidence between the pump pulse and the photons detected by Alice and Bob is computed via a time tagger with \SI{10}{\pico\second} rms jitter and \SI{2}{\nano\second} dead time.

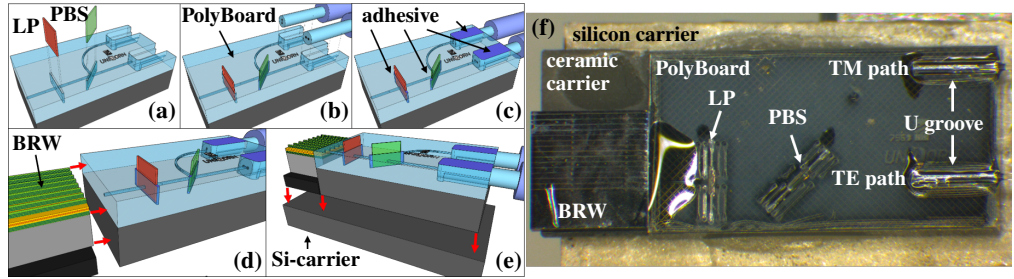
\begin{figure}[!htbp]
	\captionsetup{singlelinecheck = false, justification=raggedright}
    	 \centering
    	 \begin{annotate}{\includegraphics[width=1\textwidth]{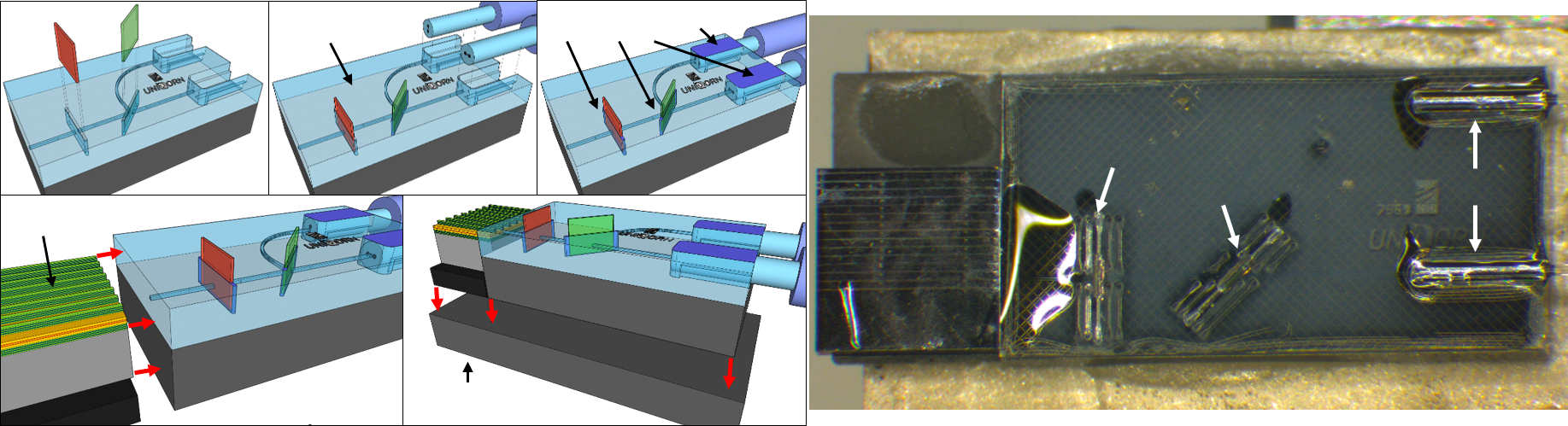}}{1}
    	\draw[very thick,black] (-6.75,1.75) node[anchor=north west]{\footnotesize\textbf{LP}};
        \draw[very thick,black] (-6.25,1.92) node[anchor=north west]{\footnotesize\textbf{PBS}};
        \draw[very thick,black] (-4.5,1.9) node[anchor=north west]{\footnotesize\textbf{PolyBoard}};
        \draw[very thick,black] (-2.1,1.9) node[anchor=north west]{\footnotesize\textbf{adhesive}};
        \draw[very thick,black] (-6.75,0.2) node[anchor=north west]{\footnotesize\textbf{BRW}};
        \draw[very thick,black] (-3.3,-1.35) node[anchor=north west]{\footnotesize\textbf{Si-carrier}};
        \draw[very thick,black] (-4.3,0.65) node[anchor=north east]{\small\textbf{(a)}};
        \draw[very thick,black] (-2,0.65) node[anchor=north east]{\small\textbf{(b)}};
        \draw[very thick,black] (0.3,0.65) node[anchor=north east]{\small\textbf{(c)}};
        \draw[very thick,black] (-3.15,-1.35) node[anchor=north east]{\small\textbf{(d)}};
        \draw[very thick,black] (0.3,-1.35) node[anchor=north east]{\small\textbf{(e)}};
        \draw[very thick,black] (0.15,1.75) node[anchor=north west]{\small\textbf{(f)}};
        \draw[very thick,white] (0.35,1.3) node[anchor=north west]{\footnotesize\textbf{ceramic}};
        \draw[very thick,white] (0.35,1.0) node[anchor=north west]{\footnotesize\textbf{carrier}};
        \draw[very thick,white] (1.8,1.2) node[anchor=north west]{\footnotesize\textbf{PolyBoard}};
        \draw[very thick,white] (0.5,-0.7) node[anchor=north west]{\footnotesize\textbf{BRW}};
        \draw[very thick,black] (0.7,1.65) node[anchor=north west]{\footnotesize\textbf{silicon carrier}};
        \draw[very thick,white] (2.5,0.8) node[anchor=north west]{\footnotesize\textbf{LP}};
        \draw[very thick,white] (3.3,0.55) node[anchor=north west]{\footnotesize\textbf{PBS}};
        \draw[very thick, white] (4.1,1.2) node[anchor=north west]{\footnotesize\textbf{TM path}};
        \draw[very thick,white] (5.1,0.47) node[anchor=north west]{\footnotesize\textbf{U groove}};
        \draw[very thick, white] (4.14,-0.26) node[anchor=north west]{\footnotesize\textbf{TE path}};
    	 \end{annotate}
        \caption{\textbf{Assembly process and photograph of the hybrid PIC.}
        First, the PolyBoard is prepared by inserting the thin-film long-pass filter (LP) and polarizing beam splitter (PBS) in their pre-etched slots~(a), installing the output fibers~(b), optimizing all elements for transmission and securing them with UV-curing, index-matched adhesive~(c).
        Next, using active alignment, the BRW is end-facet coupled to the PolyBoard and the interface secured with adhesive once the transmission is optimized~(d).
        Finally, the newly formed hybrid PIC is mechanically stabilized by a common silicon mount with 7x10\,mm footprint~(e).
        A photograph (reprinted with permission from \cite{Trenti:22}) of the final assembly used in this work~(f).}
        \label{fig:PIC_Photo}
\end{figure}

The hybrid PIC comprises a nonlinear BRW with a high $\chi^{\left(2\right)}$ nonlinear coefficient enabling efficient parametric down-conversion (PDC)~\cite{Mobini2022,Shoji2002}, and the PolyBoard, a passive optical interposer.
The assembly process and final chip are shown in Fig.~\ref{fig:PIC_Photo}.

To provide waveguiding and modal phase-matching, the BRW is made up of layers with different aluminum concentrations and etched into a ridge structure.
By carefully engineering the layer thicknesses and aluminum concentrations~\cite{Pressl2018}, and by reducing the waveguide ridge sidewall roughness~\cite{Thiel23} we achieve high photon-pair rates of up to \SI{8.9(0.5)e4}{\hertz} per mW of external pump power, which corresponds to about \SI{4e5}{\hertz} per mW of internal pump power~\cite{Nardi2022}.
The photon pairs generated in the telecom wavelength range benefit from minimal signal attenuation in the existing fiber infrastructure.
Because of their broad-band ($\sim$\SI{100}{\nano\meter}) emission, BRWs can also be considered for the distribution of entanglement in multiple telecom channels.
In addition to being correlated in their time of creation, which is used for time-bin entanglement, the two photons of a pair are anti-correlated in wavelength and polarization, opening up the possibility for other forms of entanglement or even hyperentanglement.
Achievements realized with BRWs thus far include the generation of polarization entanglement~\cite{Horn2013,Valles2013,Schlager2017}, energy-time entanglement~\cite{Autebert2016}, and free-space time-bin entanglement~\cite{Chen2018}, as well as the integration of an internal pump laser~\cite{Bijlani2013,Boitier2014} and with it the demonstration of difference-frequency generation~\cite{Schlager2021}.
\\
The photons generated by PDC in the BRW are orthogonally polarized and collinear with the pump light.
It is therefore necessary to spectrally filter the pump and convenient to separate the photons with a polarizing beamsplitter.
The required components are technologically challenging to realize and therefore we employ a hybrid integration with polymer waveguide circuits~\cite{Kleinert:15}.
\\

Polymer-based PICs feature lower production and material cost than standard semiconductor platforms~\cite{Rahlves:15,Rezem:17}, a large transparency window, and an effective index that closely matches silica fibers allowing low-loss pigtailing.
The presented interposer features two custom-made, dielectric thin-film elements, a LP to reject the pump light and a polarizing beam splitter (PBS) redirecting orthogonal polarizations to separate waveguides.
The input waveguide facet is diced for end-facet coupling to the BRW, whereas the output waveguides are directly pigtailed with standard polarization maintaining (PM) fibers in a U-groove arrangement~\cite{Kleinert:15}, which improves the mechanical stability.

Hybrid integration, as employed between the BRW and the PolyBoard in this work, combines the strengths of both material platforms and can also introduce new features missing in the monolithic counterparts.
The PolyBoard has proven its versatility by implementing on-chip free space sections~\cite{Kleinert:17}, thermal phase shifters or switches~\cite{Zhang:16}, tunable distributed Bragg-reflector lasers~\cite{deFelipe:14, Happach:17}, on-chip isolators and circulators~\cite{Conradi:2018, 10.1117/12.2545592                              }, and various integrated circuits for quantum photonics~\cite{Trenti:22}.

When interfacing dissimilar platforms, the mode field overlap is crucial for the coupling loss.
The complicated layer structure of the BRW gives rise to a non-rotationally symmetric mode with a shape resembling two stacked cigars.
Thus, the current design results in a mode field overlap with the near-Gaussian mode of the PolyBoard of~$\sim$\SI{55}{\percent} (for more details see Appendix~\ref{app:PolyBoard} and~\ref{app:Modes_Loss}) but mode-engineering via taper structures can boost the overlap significantly.
The assembly of the hybrid PIC is a multi-step process with active alignment using the telecom laser transmission signal and is sketched in Fig.~\ref{fig:PIC_Photo}~(a)-(e).

\section{Results\label{sec:results}}
We perform a series of classical characterization measurements to evaluate the performance of the individual components of the PIC.
The results provide insights in addition to the coincidence measurements at the few-photon level and, furthermore, are less sensitive to noise. 

\begin{figure}[!htbp]
	\captionsetup{singlelinecheck = false, justification=raggedright}
    	 \centering
    	 \begin{annotate}{\includegraphics[width=\textwidth]{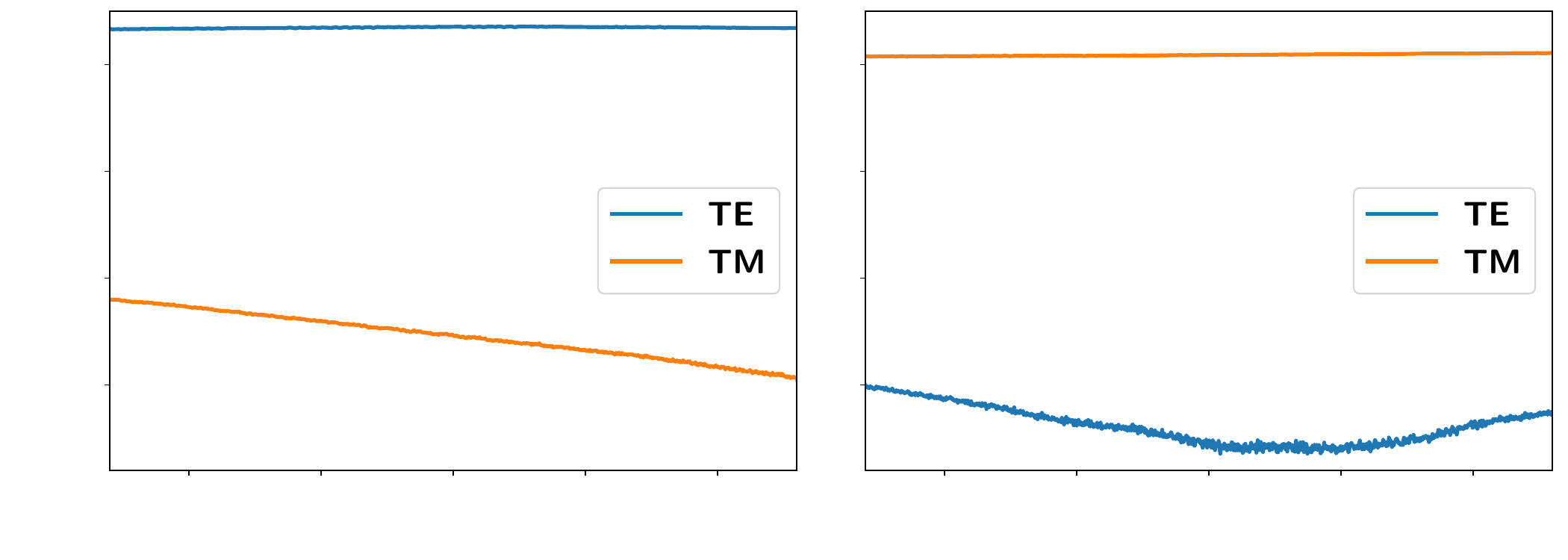}}{1}
        \draw[very thick,black] (-5.8, 1.8) node[anchor=north west]{\textbf{(a)}};
        \draw[very thick,black] (0.63,1.8) node[anchor=north west]{\textbf{(b)}};
    	\draw[very thick,black] (-4.,-2.25) node[anchor=north west]{Wavelength (nm)};
        \draw[very thick,black] (-5.55,-1.7) node[anchor=north west]{1522};
        \draw[very thick,black] (-3.35,-1.7) node[anchor=north west]{1528};
        \draw[very thick,black] (-1.1,-1.7) node[anchor=north west]{1534};
        \draw[very thick,black] (2.5,-2.25) node[anchor=north west]{Wavelength (nm)};
        \draw[very thick,black] (0.9,-1.7) node[anchor=north west]{1522};
        \draw[very thick,black] (3.1,-1.7) node[anchor=north west]{1528};
        \draw[very thick,black] (5.35,-1.7) node[anchor=north west]{1534};
        \draw[very thick,black] (-7,-1.2) node[anchor=north west, rotate=90]{Transmission (dB)};
        \draw[very thick,black] (-6.45,2.1) node[anchor=north west]{-10};
        \draw[very thick,black] (-6.45,1.2) node[anchor=north west]{-20};
        \draw[very thick,black] (-6.45,0.3) node[anchor=north west]{-30};
        \draw[very thick,black] (-6.45,-0.65) node[anchor=north west]{-40};
         \draw[very thick,black] (0,2.1) node[anchor=north west]{-10};
        \draw[very thick,black] (0,1.2) node[anchor=north west]{-20};
        \draw[very thick,black] (0,0.3) node[anchor=north west]{-30};
        \draw[very thick,black] (0,-0.65) node[anchor=north west]{-40};
    	 \end{annotate}
        \caption{\textbf{Transmission measurements of the PolyBoard interposer.} The input polarization is set to transversal-electric (TE) and transversal-magnetic (TM) and the transmission is measured at both output paths while the laser wavelength is scanned. The polarizing beam splitter (PBS) predominantly transmits TE-polarized light and reflects TM-polarized light. For both the TE path (a) and the TM path (b), the polarization extinction ratio is calculated as the difference between TE and TM transmission. The spectral dependence of the suppressed polarization is ascribed to chromatic effects in the PBS thin-film layer stack.}
        \label{fig:Classical_Meas}
\end{figure}
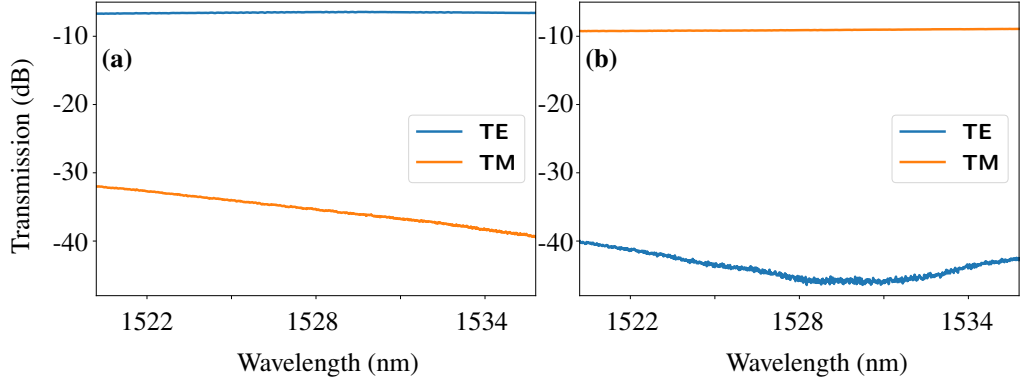

To this end, we couple a CW laser into the diced facet of the PolyBoard and measure the transmission of both output fibers for transversal-electrically (TE) and transversal-magnetically (TM) polarized input light while scanning the laser wavelength (see Fig.~\ref{fig:Classical_Meas}).
For both outputs, we find a flat transmission curve for the favored polarization with an average loss of \SI{6.54(8)}{\deci\bel} for the TE and \SI{9.1(1)}{\deci\bel} for the TM path. Note that this measurement also includes the input coupling loss of \SIrange{0.5}{1}{\deci\bel}.
The lower transmission of the TM fiber is ascribed to a slight out-of-plane deflection caused by a non-optimal angle of the inserted PBS.
Further, we compute the polarization extinction ratios from the transmission in the orthogonal polarization yielding $P\!E\!R>\SI{30}{\deci\bel}$ for the reflection and $P\!E\!R>\SI{25}{\deci\bel}$ for the transmission port of the PBS.

The suppression of pump light by the LP cannot be measured directly at the PolyBoard but is inferred from a separate test structure. 
Using a white light source, we find a suppression exceeding \SI{40}{\deci\bel} for the range of \SIrange{700}{850}{\nano\meter} limited only by the noise floor of our detector. Employing a laser diode emitting at \SI{785}{\nano\meter}, we verify a suppression of \SI{68(1)}{\deci\bel}, while the loss of the LP at telecom wavelengths amounts to $\sim\SI{0.9}{\deci\bel}$ (for more details see Appendix~\ref{app:PolyBoard}).

We conclude that the PolyBoard not only reduces size and cost of the implementation drastically but also provides high-performance polarization splitting and long-pass filtering that easily matches or even outperforms the characteristics of bulk elements.

Moving on to characterizing the quantum performance of our PIC, we pump PDC by coupling 767\,nm CW light into the BRW input facet.
We measure a coincidence rate of 460\,Hz per mW external pump power between the signal and idler photons without correcting for fiber loss or detector efficiency.
The coincidence rate is consistent with this BRW's stand-alone performance considering the losses expected from hybrid integration on the PIC described above.

For the time-bin entanglement measurement and state tomography, we follow the methods described by James et al.~\cite{James2001} and Takesue et al.~\cite{Takesue2009}.
We reference the arrival times of photons at Alice's and Bob's detectors to a trigger given by a photodiode installed in the path of the pulsed pump laser.
Due to the limited transmissions of the free-space interferometers of \SIrange{5}{7}{\%}, we measure a total coincidence count of about 1.4\,Hz per mW external pump power.
Correcting for the loss in the two telecom interferometers we obtain a coincidence rate of \SIrange{290}{560}{\Hz}.
By rotating the phase plate in one of the interferomters, we reveal the interference in the central time bin with a \SI{91(5)}{\percent} visibility.

From the triple coincidence between the trigger and Alice's and Bob's detectors results a 2D histogram, an example of which is shown in Appendix~\ref{app:Hist}.
We perform a measurement for each of the four states $\ket{++}$, $\ket{+L}$, $\ket{L+}$, and $\ket{LL}$, where $\ket{+}=1/\sqrt{2}\left(\ket{1}+\ket{2}\right)$ and $\ket{L}=1/\sqrt{2}\left(\ket{1}+i\ket{2}\right)$, by rotating the phase plates in Alice's and Bob's interferometers.
From these, we obtain the coincidence counts (without correcting for accidentals) for projections onto 16 different two-photon states serving as input for the state tomography.
As the linear reconstruction of the density matrix leads to negative eigenvalues and therefore an unphysical state, we employ a maximum likelihood estimation to recover the density matrix shown in Fig.~\ref{fig:Density_Matrix} (values can be found in Appendix~\ref{app:matrix}).
We obtain a concurrence of $\left(96_{-8}^{+3}\right)\%$, a $\left(96_{-5}^{+2}\right)\%$ fidelity to the $\ket{\Phi^+}$ Bell state and a Bell S-parameter of $\left(2.70_{-0.33}^{+0.09}\right)\%$.
The uncertainties are derived using a Monte Carlo simulation where we create $10^4$ sets of coincidence counts with Poissonian distribution around the actually measured counts and perform the maximum likelihood estimation for each.
The results demonstrate both strong entanglement and violation of the Clauser-Horne-Shimony-Holt (CHSH) Bell inequality.
The resulting nonlocal correlations are a useful resource for quantum communication tasks.
    
\begin{figure}[!htbp]
	\captionsetup{singlelinecheck = false, justification=raggedright}
    	 \centering
    	 \begin{annotate}{\includegraphics[width=0.8\textwidth]{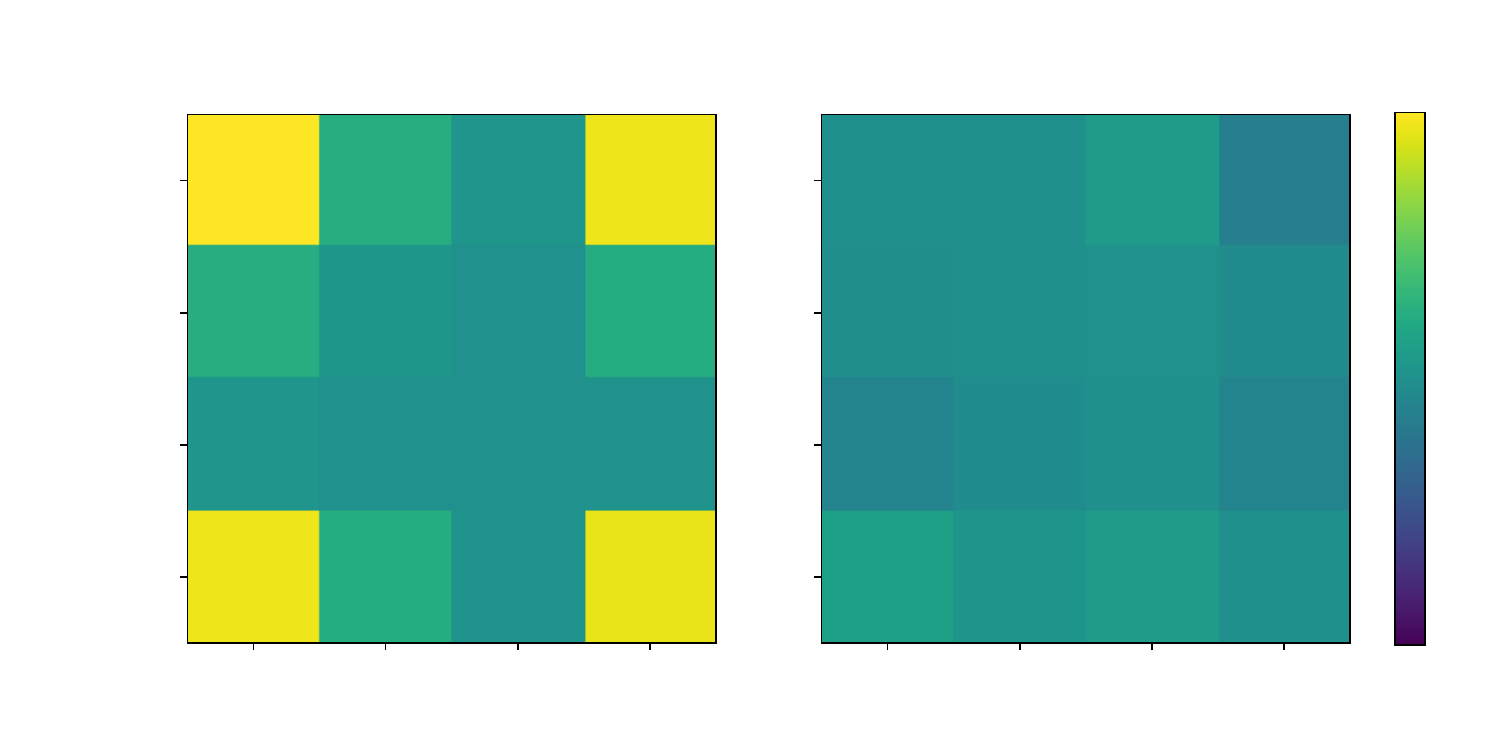}}{1}
    	\draw[very thick,black] (-3,2.3) node[anchor=north west]{Real Part};
        \draw[very thick,black] (1.1,2.3) node[anchor=north west]{Imaginary Part};
        \draw[very thick,black] (-4.8,1.7) node[anchor=north west]{$\ket{11}$};
        \draw[very thick,black] (-4.8,0.7) node[anchor=north west]{$\ket{12}$};
        \draw[very thick,black] (-4.8,-0.3) node[anchor=north west]{$\ket{21}$};
        \draw[very thick,black] (-4.8,-1.3) node[anchor=north west]{$\ket{22}$};
        \draw[very thick,black] (-3.9,-2.1) node[anchor=north west]{$\ket{11}$};
        \draw[very thick,black] (-3,-2.1) node[anchor=north west]{$\ket{12}$};
        \draw[very thick,black] (-2.1,-2.1) node[anchor=north west]{$\ket{21}$};
        \draw[very thick,black] (-1.2,-2.1) node[anchor=north west]{$\ket{22}$};
        \draw[very thick,black] (-0.3,1.7) node[anchor=north west]{$\ket{11}$};
        \draw[very thick,black] (-0.3,0.7) node[anchor=north west]{$\ket{12}$};
        \draw[very thick,black] (-0.3,-0.3) node[anchor=north west]{$\ket{21}$};
        \draw[very thick,black] (-0.3,-1.3) node[anchor=north west]{$\ket{22}$};
        \draw[very thick,black] (0.6,-2.1) node[anchor=north west]{$\ket{11}$};
        \draw[very thick,black] (1.5,-2.1) node[anchor=north west]{$\ket{12}$};
        \draw[very thick,black] (2.5,-2.1) node[anchor=north west]{$\ket{21}$};
        \draw[very thick,black] (3.5,-2.1) node[anchor=north west]{$\ket{22}$};
        \draw[very thick,black] (5,2.2) node[anchor=north west]{0.5};
        \draw[very thick,black] (5,-1.6) node[anchor=north west]{-0.5};
        \draw[very thick,black] (4.67,2) node[anchor=north west]{-};
        \draw[very thick,black] (4.67,-1.78) node[anchor=north west]{-};
        \draw[very thick,black] (4.67,0.11) node[anchor=north west]{-};
        \draw[very thick,black] (5,0.2) node[anchor=north west]{0.0};
    	 \end{annotate}
        \caption{\textbf{Density matrix reconstructed via maximum likelihood estimation.} The real and imaginary parts of the density matrix demonstrate the high degree of entanglement and fidelity to the $\ket{\Phi^+}$ Bell state.}
        \label{fig:Density_Matrix}
\end{figure}
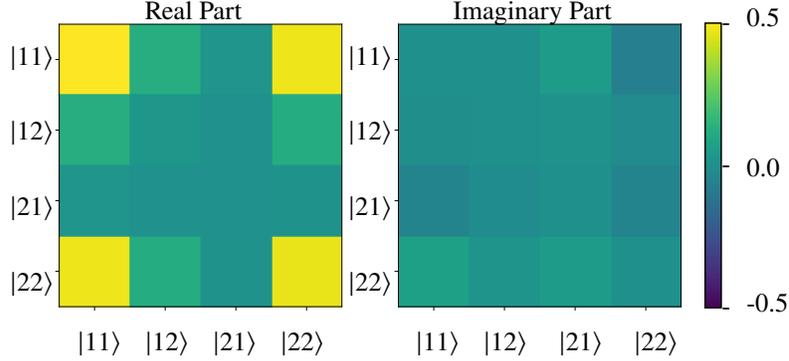

\section{Discussion and conclusion}
The mass-deployment of entanglement-based QKD transceivers requires a high level of integration while components must comply with the challenging operation in a real-life environment based on noisy, dispersive fiber networks.
The pair emission rate of our PIC is consistent with previous experiments using BRWs in our group~\cite{Chen2018} and comparable to those of others~\cite{Appas2022}.
Further enhancement of the coincidence rates involves optimizing the design of the hybrid PIC.
First, engineering the BRW's and the PolyBoard's mode field at the intersection using tapered waveguides reduces loss due to mode mismatch.
Second, employing the latest generation of BRWs -- featuring photon-pair generation rates >60 times higher than the sample used here~\cite{Nardi2022} -- significantly relaxes the requirements for low loss down the line.
In the current implementation, we identify the free-space interferometers as the dominating source of loss and therefore as the bottleneck on our way to efficiently produce entangled photon pairs.
Actively stabilizing the interferometers can improve the spatial overlap of beams as well as the temporal overlap of pulses and counteract some of the degradation in the classical visibilities~\cite{Toliver2015}.
However, the free-space interferometers are not at the core of this work.
Exchanging them for chip- or fiber-based counterparts may not only improve the efficiency but also promotes miniaturisation further.

In contrast to the modest coincidence rates, the demonstrated entanglement is very strong with a concurrence of $\left(96_{-8}^{+3}\right)\%$ and a fidelity of $\left(96_{-5}^{+2}\right)\%$ to a Bell state.
Lower uncertainties can be achieved with higher count rates once the interferometers have been replaced.
Already now, our PIC compares well with other telecom time-bin entanglement demonstrations, including the \SI{88.9(1.8)}{\percent} concurrence and \SI{94.2(9)}{\percent} fidelity measured for a bare BRW~\cite{Chen2018} in free space, the \SI{74.1(4.8)}{\percent} coincidence fringe visibility found for a fiber-based approach~\cite{Takesue2014}, and the \SI{91.0(7)}{\percent} fidelity quoted for an all on-chip implementation~\cite{Zhang2018}.

We attribute the increased purity of the entanglement to the hybrid integration of BRW and PolyBoard, as the PIC offers optical stability and the end-facet coupling reduces the amount of unwanted photoluminescence picked up from the BRW~\cite{Auchter2021}.
Moreover, the $P\!E\!R$ of $>\SI{25}{\deci\bel}$ reduces the rate of the accidental coincidences and the strong suppression of the LP of \SI{68(1)}{\deci\bel} obviates the need for additional bandwidth filtering or background suppression.
By adding thin-film elements for band-pass filtering or chromatic pre-compensation, we expect to reduce effects of dispersion and thereby improve the temporal overlap of the pulses.
Finally, balancing the loss of both polarization modes will enhance the entanglement further.

To conclude, we demonstrated the hybrid integration of a BRW with the PolyBoard interposer to produce high-quality time-bin entangled photon pairs in the telecom wavelength range.
Our results testify to the adequacy of the BRW-PolyBoard PIC for miniaturized quantum communication.
We identify the main causes of photon loss and outline a feasible route towards a second generation of significantly enhanced hybrid PICs.
Here, the most notable upgrades include transitioning to fiber- or chip-based interferometers, engineering the mode field overlap at the chip interface and employing the already available and greatly improved BRW structures.

\begin{appendices}
\section{PolyBoard\label{app:PolyBoard}}
The PolyBoard is fabricated from two different polymers which are iteratively applied via spin-coating on a 4-inch silicon wafer and further processed using photo-lithography and dry-etching.
The waveguides have a quadratic cross-section of $3.2\,\mu$m\,x\,$3.2\,\mu$m with an index contrast of $\Delta n=0.03$ resulting in a simulated mode field that is rotationally symmetric with a $1/e^2$ diameter of $3.9\,\mu$m.
The effective index for the transversal-electric (TE) mode is $n_{\text{eff}}^{\text{TE}}=$\SI{1.463} whereas the transversal-magnetic (TM) mode has $n_{\text{eff}}^{\text{TM}}=$\SI{1.462} resulting in a birefringence of $\sim\SI{1E-3}{}$.
Using cut-back measurements, we evaluate a propagation loss of $\sim\SI{0.9}{\deci\bel\per\centi\meter}$ for this wafer in separate test structures.

In the presented PolyBoard, thin-film elements (TFE) are used to realize wavelength filtering and polarization splitting which are challenging to implement monolithically as they require a large footprint or exhibit high losses and low extinction ratios.
During the fabrication, slots of a few-micrometer thickness are etched into the PolyBoard and are later equipped with the fitting TFE~\cite{Kleinert:17}.
Because of the small index contrast, the optical loss caused by the unguided propagation through the etched slot is limited and further minimized by appropriate waveguide tapering on each side of the slot.
After inserting the TFEs, they are secured with an index-matched and UV-curable adhesive.

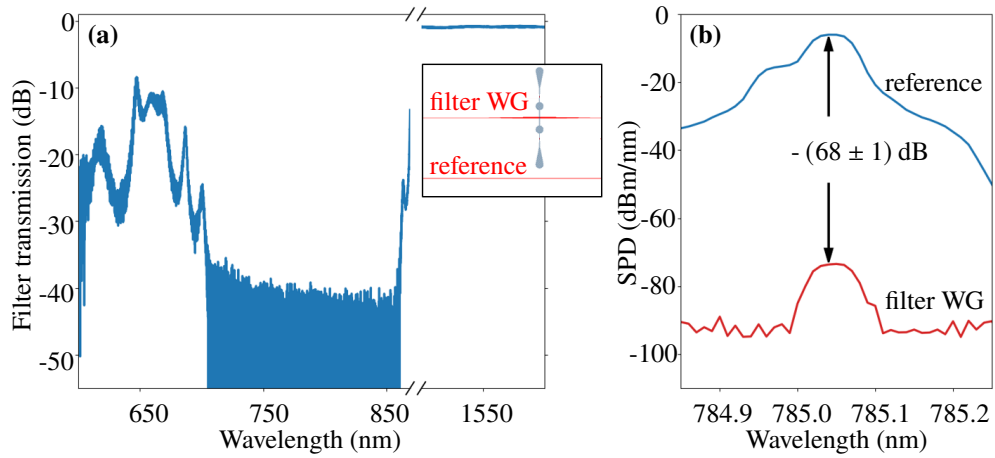
\begin{figure}[!htbp]
	\captionsetup{singlelinecheck = false, justification=raggedright}
    	 \centering
    	 \begin{annotate}{\includegraphics[width=\textwidth]{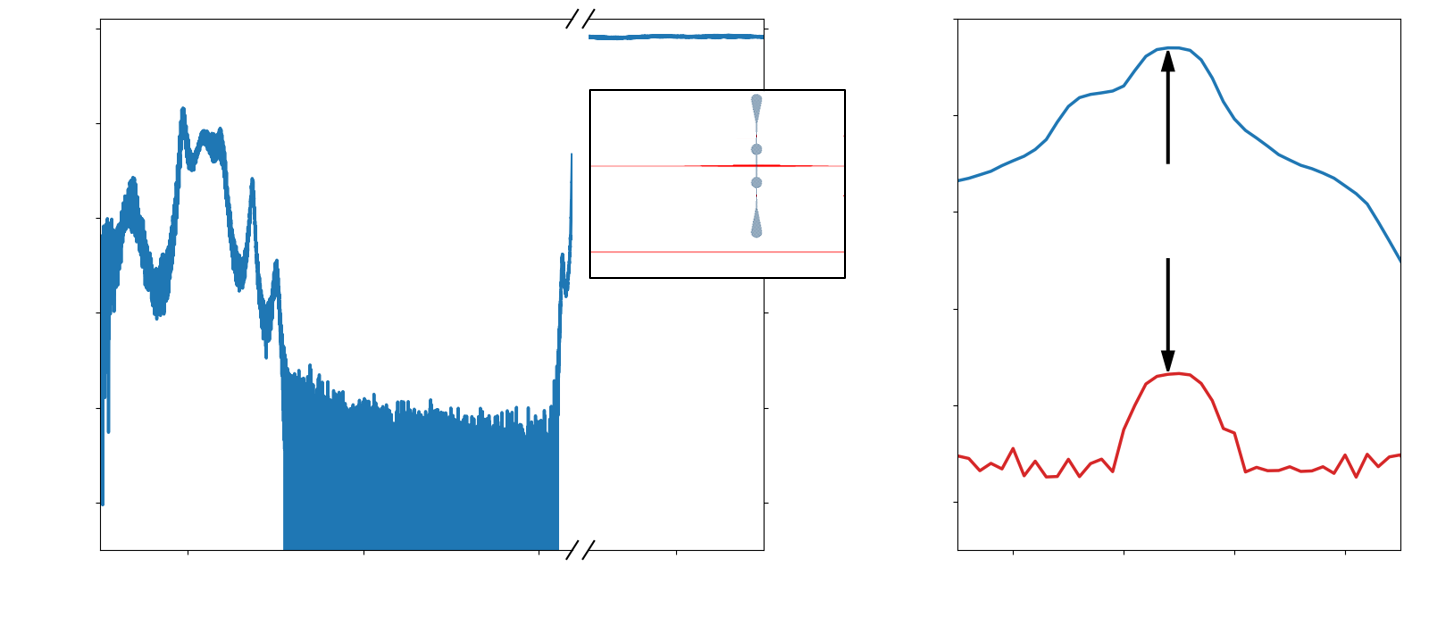}}{1}
    	\draw[very thick,black] (-4,-2.65) node[anchor=north west]{Wavelength (nm)};
        \draw[very thick,black] (-6.75,-1.6) node[anchor=north west, rotate=90]{Filter transmission (dB)};
        \draw[very thick,black] (-5.75,2.75) node[anchor=north west]{\textbf{(a)}};
        \draw[very thick,black] (-6.1,2.9) node[anchor=north west]{0};
        \draw[very thick,black] (-6.4,2) node[anchor=north west]{-10};
        \draw[very thick,black] (-6.4,1.1) node[anchor=north west]{-20};
        \draw[very thick,black] (-6.4,0.2) node[anchor=north west]{-30};
        \draw[very thick,black] (-6.4,-0.7) node[anchor=north west]{-40};
        \draw[very thick,black] (-6.4,-1.6) node[anchor=north west]{-50};
        \draw[very thick,black] (-5.2,-2.3) node[anchor=north west]{650};
        \draw[very thick,black] (-3.6,-2.3) node[anchor=north west]{750};
        \draw[very thick,black] (-2.0,-2.3) node[anchor=north west]{850};
        \draw[very thick,black] (-0.8,-2.3) node[anchor=north west]{1550};
        \draw[very thick,red] (-1.2,1.85) node[anchor=north west]{filter WG};
        \draw[very thick,red] (-1.2,1) node[anchor=north west]{reference};
        \draw[very thick,black] (3,-2.65) node[anchor=north west]{Wavelength (nm)};
        \draw[very thick,black] (1.3,-1) node[anchor=north west, rotate=90]{SPD (dBm/nm)};
        \draw[very thick,black] (4.85,-0.8) node[anchor=north west]{filter WG};
        \draw[very thick,black] (4.85,2.1) node[anchor=north west]{reference};
        \draw[very thick,black] (3.6,1.2) node[anchor=north west]{-\,\SI{68(1)}{\deci\bel}};
        \draw[very thick,black] (2.25,2.75) node[anchor=north west]{\textbf{(b)}};
        \draw[very thick,black] (1.8,3) node[anchor=north west]{0};
        \draw[very thick,black] (1.6,2.1) node[anchor=north west]{-20};
        \draw[very thick,black] (1.6,1.2) node[anchor=north west]{-40};
        \draw[very thick,black] (1.6,0.3) node[anchor=north west]{-60};
        \draw[very thick,black] (1.6,-0.6) node[anchor=north west]{-80};
        \draw[very thick,black] (1.45,-1.5) node[anchor=north west]{-100};
        \draw[very thick,black] (2.3,-2.3) node[anchor=north west]{784.9};
        \draw[very thick,black] (3.35,-2.3) node[anchor=north west]{785.0};
        \draw[very thick,black] (4.4,-2.3) node[anchor=north west]{785.1};
        \draw[very thick,black] (5.45,-2.3) node[anchor=north west]{785.2};
    	 \end{annotate}
        \caption{\textbf{Transmission measurements for long-pass (LP) filter at test structure.} Filter transmission over a broad spectrum using a white light source and a tunable C-Band laser (a) and a laser diode emitting at \SI{785.05}{\nano\meter} with high spectral power density ($S\!P\!D$) (b). The filter performance is extracted from the transmission of a straight reference waveguide (WG) and a waveguide passing the filter slot with the LP. The layout of the test structure is shown in the inset of (a).}
        \label{fig:Filter}
\end{figure}
To assess the performance of the long-pass (LP) filter, we insert it into a test structure depicted in the inset of Fig.~\ref{fig:Filter} (a).
By comparing the transmission of a straight reference waveguide and a waveguide passing the LP, we evaluate the suppression. 
We perform the measurement for three different light sources: a supercontinuum white light laser (\textit{NKT Photonics, SuperK}), a laser diode emitting at \SI{785.05}{\nano\meter} with high spectral power density (\textit{Integrated optics, Matchbox 785nm SLM}), and a tunable laser (\textit{Agilent, 8164B with 81635A and 81689B}) covering the telecom C-band.
In this way, we analyze the transmission for telecom wavelengths, the maximum suppression close to the wavelength used to pump the parametric down-conversion process, and the bandwidth of the filter. We find a suppression of more than $\SI{40}{\deci\bel}$ for \SIrange{700}{850}{\nano\meter} and a maximum suppression of \SI{68(1)}{\deci\bel} at \SI{785.05}{\nano\meter}.

\section{Modes and coupling loss\label{app:Modes_Loss}}
\begin{figure}[!htbp]
	\captionsetup{singlelinecheck = false, justification=raggedright}
    	 \centering
    	 \begin{annotate}{\includegraphics[width=0.8\textwidth]{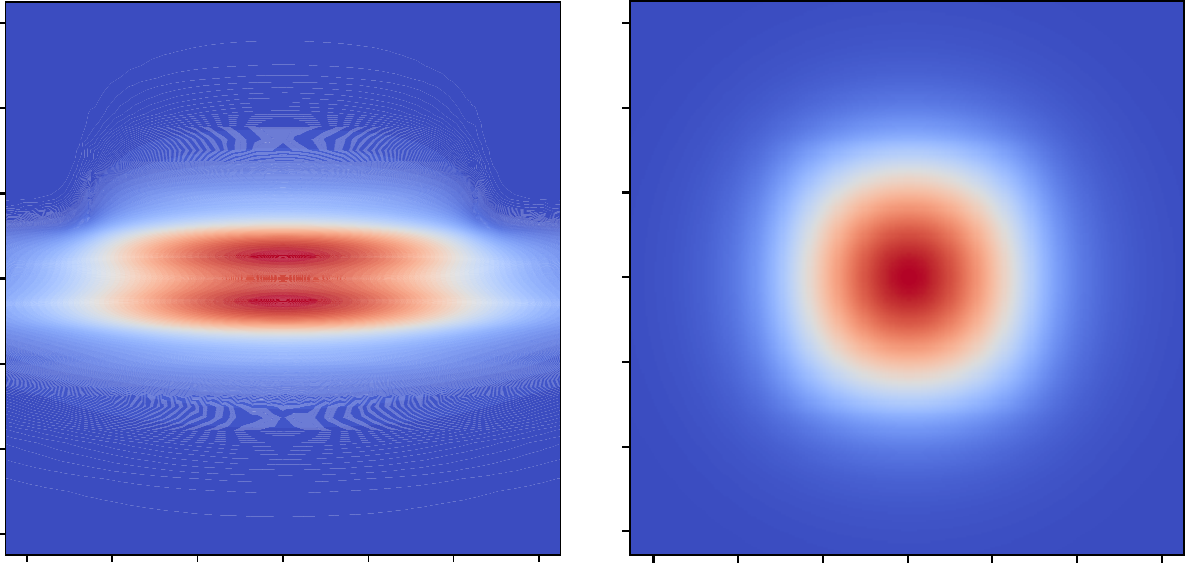}}{1}
        \draw[very thick,black] (-6.05,2.93) node[anchor=north west]{\textbf{(a)}};
        \draw[very thick,black] (-6.35,-1.7) node[anchor=north west, rotate=90]{Vertical Direction (\SI{}{\micro\meter})};
        \draw[very thick,black] (-4.85,-2.9) node[anchor=north west]{Horizontal Direction (\SI{}{\micro\meter})};
        \draw[very thick,black] (1.05,-2.9) node[anchor=north west]{Horizontal Direction (\SI{}{\micro\meter})};
        \draw[very thick,black] (-5.73,2.62) node[anchor=north west]{3};
        \draw[very thick,black] (-5.73,1.85) node[anchor=north west]{2};
        \draw[very thick,black] (-5.73,1.05) node[anchor=north west]{1};
        \draw[very thick,black] (-5.73,0.25) node[anchor=north west]{0};
        \draw[very thick,black] (-5.75,-0.52) node[anchor=north west]{-1};
        \draw[very thick,black] (-5.75,-1.28) node[anchor=north west]{-2};
        \draw[very thick,black] (-5.75,-2) node[anchor=north west]{-3};
        \draw[very thick,black] (-5.4,-2.5) node[anchor=north west]{-3};
        \draw[very thick,black] (-4.65,-2.5) node[anchor=north west]{-2};
        \draw[very thick,black] (-3.76,-2.5) node[anchor=north west]{-1};
        \draw[very thick,black] (-2.9,-2.5) node[anchor=north west]{0};
        \draw[very thick,black] (-2.11,-2.5) node[anchor=north west]{1};
        \draw[very thick,black] (-1.42,-2.5) node[anchor=north west]{2};
        \draw[very thick,black] (-0.7,-2.5) node[anchor=north west]{3};
        \draw[very thick,black] (-0.3,2.93) node[anchor=north west]{\textbf{(b)}};
        \draw[very thick,black] (-0.15,2.62) node[anchor=north west]{3};
        \draw[very thick,black] (-0.15,1.85) node[anchor=north west]{2};
        \draw[very thick,black] (-0.15,1.05) node[anchor=north west]{1};
        \draw[very thick,black] (-0.15,0.25) node[anchor=north west]{0};
        \draw[very thick,black] (-0.17,-0.52) node[anchor=north west]{-1};
        \draw[very thick,black] (-0.17,-1.28) node[anchor=north west]{-2};
        \draw[very thick,black] (-0.17,-2) node[anchor=north west]{-3};
        \draw[very thick,black] (0.3,-2.5) node[anchor=north west]{-3};
        \draw[very thick,black] (1.12,-2.5) node[anchor=north west]{-2};
        \draw[very thick,black] (1.92,-2.5) node[anchor=north west]{-1};
        \draw[very thick,black] (2.7,-2.5) node[anchor=north west]{0};
        \draw[very thick,black] (3.5,-2.5) node[anchor=north west]{1};
        \draw[very thick,black] (4.16,-2.5) node[anchor=north west]{2};
        \draw[very thick,black] (4.94,-2.5) node[anchor=north west]{3};
    	 \end{annotate}
        \caption{\textbf{Simulations of the waveguide modes of BRW and PolyBoard.} The electric field absolute value of the 1550\,nm TE-polarized double cigar shaped mode of the BRW (a) and the rotationally symmetric mode of the PolyBoard waveguide (b). The TM-polarized modes are not shown here as they look very similar.}
        \label{fig:Modes}
\end{figure}

\begin{figure}[!htbp]
	\captionsetup{singlelinecheck = false, justification=raggedright}
    	 \centering
        \begin{annotate}{\includegraphics[width=0.8\textwidth]{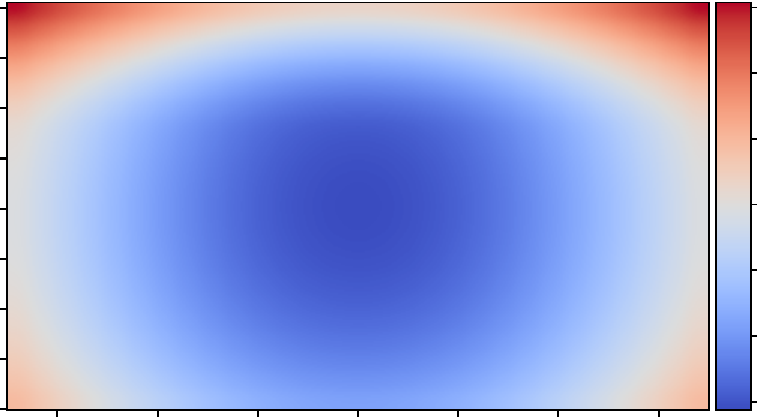}}{1}
        \draw[very thick,black] (-2.6,-3.4) node[anchor=north west]{Horizontal Displacement (\SI{}{\micro\meter})};
        \draw[very thick,black] (-6.3,-2) node[anchor=north west, rotate=90]{Vertical Displacement (\SI{}{\micro\meter})};
        \draw[very thick,black] (-5.8,3) node[anchor=north west]{2};
        \draw[very thick,black] (-5.8,1.6) node[anchor=north west]{1};
        \draw[very thick,black] (-5.8,0.25) node[anchor=north west]{0};
        \draw[very thick,black] (-5.82,-1.18) node[anchor=north west]{-1};
        \draw[very thick,black] (-5.82,-2.6) node[anchor=north west]{-2};
        \draw[very thick,black] (-4.85,-2.9) node[anchor=north west]{-3};
        \draw[very thick,black] (-3.32,-2.9) node[anchor=north west]{-2};
        \draw[very thick,black] (-1.95,-2.9) node[anchor=north west]{-1};
        \draw[very thick,black] (-0.5,-2.9) node[anchor=north west]{0};
        \draw[very thick,black] (0.92,-2.9) node[anchor=north west]{1};
        \draw[very thick,black] (2.33,-2.9) node[anchor=north west]{2};
        \draw[very thick,black] (3.75,-2.9) node[anchor=north west]{3};
        \draw[very thick,black] (5.9,-1.2) node[anchor=north west, rotate=90]{Coupling Loss (dB)};
        \draw[very thick,black] (5.35,3.09) node[anchor=north west]{21};
        \draw[very thick,black] (5.35,2.15) node[anchor=north west]{18};
        \draw[very thick,black] (5.35,1.23) node[anchor=north west]{15};
        \draw[very thick,black] (5.35,0.3) node[anchor=north west]{12};
        \draw[very thick,black] (5.35,-0.6) node[anchor=north west]{9};
        \draw[very thick,black] (5.35,-1.55) node[anchor=north west]{6};
        \draw[very thick,black] (5.35,-2.48) node[anchor=north west]{3};
        \end{annotate}
        \caption{\textbf{Coupling loss between BRW and PolyBoard waveguides.} The minimum achievable coupling loss is~$\sim$\SI{2.6}{\dB}. Due to the layer structure of the BRW, even a small displacement in the vertical direction leads to a significant increase in coupling loss.}
        \label{fig:CouplingLoss}
\end{figure}
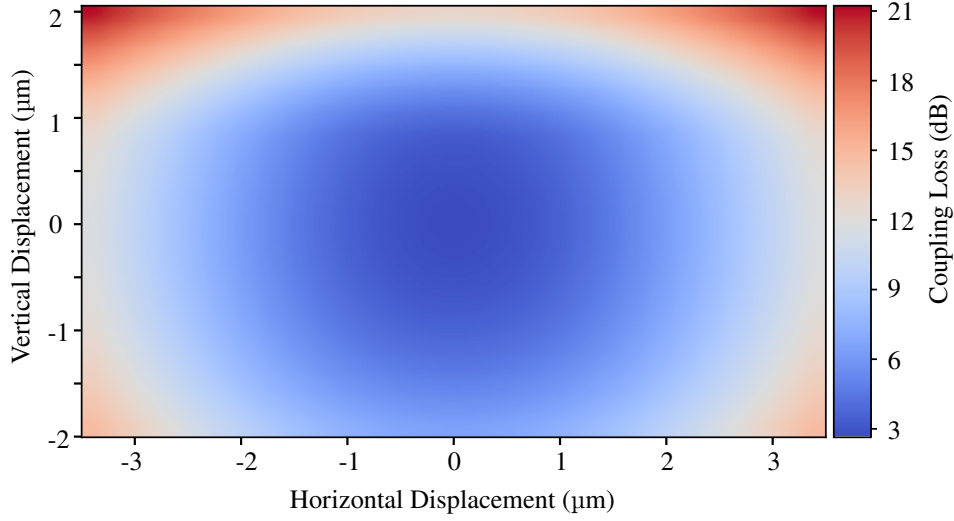

The telecom wavelength optical modes of the Bragg-reflection waveguide (BRW) and the PolyBoard have very different shapes, as shown in Fig.~\ref{fig:Modes}. Upon assembly of the photonic integrated circuit, any displacement of the facets with respect to each other leads to a considerable coupling loss, as shown in Fig.~\ref{fig:CouplingLoss}. For perfect alignment we expect $\sim\SI{2.6}{\deci\bel}$ of coupling loss with a horizontal (vertical) tolerance of $\sim\SI{1.1}{\micro\meter}$ ($\sim\SI{0.7}{\micro\meter}$) causing additional \SI{1}{\deci\bel} of loss or $\sim\SI{1.8}{\micro\meter}$ ($\sim\SI{1.3}{\micro\meter}$) causing \SI{3}{\deci\bel}.

\section{2D histogram and its interpretation\label{app:Hist}}

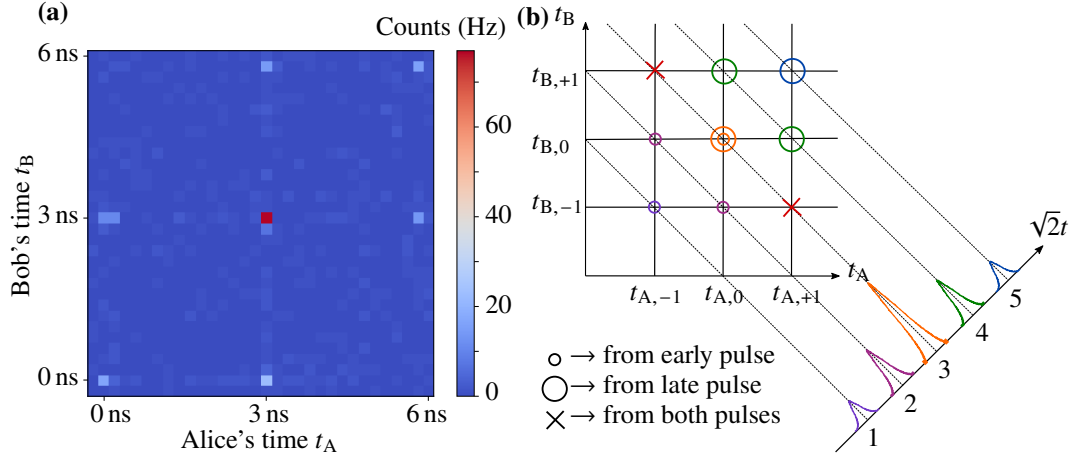
\begin{figure}[!htbp]
    \captionsetup{singlelinecheck = false, justification=raggedright}
        \begin{subfigure}[t]{0.5\textwidth}
            \centering
        	 \begin{annotate}{\includegraphics[width=\textwidth]{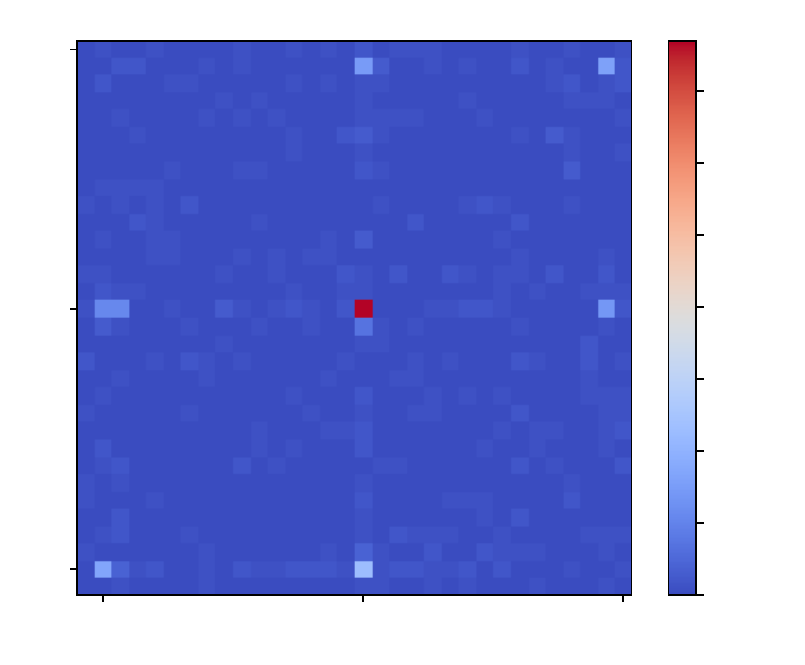}}{1}
            \draw[very thick,black] (-3.5,3.2) node[anchor=north west]{\textbf{(a)}};
            \draw[very thick,black] (-1.6,-2.5) node[anchor=north west]{Alice's time $t_\mathrm{A}$};
            \draw[very thick,black] (-3.85,-1) node[anchor=north west, rotate=90]{Bob's time $t_\mathrm{B}$};
            \draw[very thick,black] (1,3) node[anchor=north west]{Counts (Hz)};
    		\draw[very thick,black] (-2.8,-2.15) node[anchor=north west]{0\,ns};
            \draw[very thick,black] (-0.7,-2.15) node[anchor=north west]{3\,ns};
            \draw[very thick,black] (1.4,-2.15) node[anchor=north west]{6\,ns};
            \draw[very thick,black] (-3.5,-1.65) node[anchor=north west]{0\,ns};
            \draw[very thick,black] (-3.5,0.5) node[anchor=north west]{3\,ns};
            \draw[very thick,black] (-3.5,2.65) node[anchor=north west]{6\,ns};
            \draw[very thick,black] (2.45,-1.85) node[anchor=north west]{0};
            \draw[very thick,black] (2.45,-0.7) node[anchor=north west]{20};
            \draw[very thick,black] (2.45,0.5) node[anchor=north west]{40};
            \draw[very thick,black] (2.45,1.7) node[anchor=north west]{60};
    	   \end{annotate}   
        \end{subfigure}
        \begin{subfigure}[t]{0.5\textwidth}
            \centering
        	 \begin{annotate}{\includegraphics[width=\textwidth]{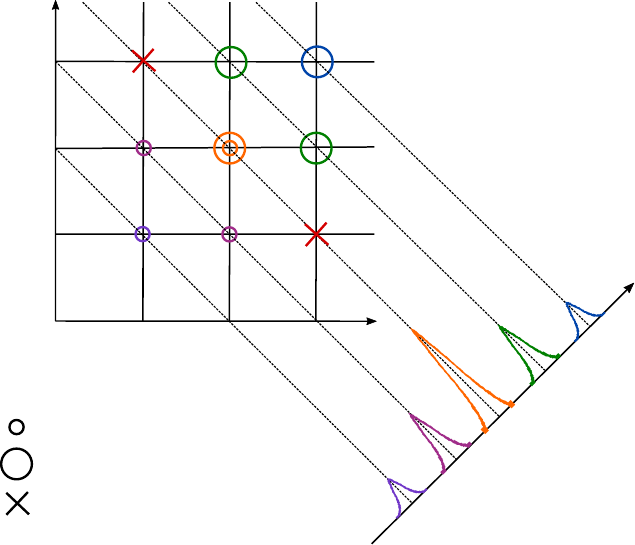}}{1}
            \draw[very thick,black] (-3.8,3.2) node[anchor=north west]{\textbf{(b)}};
            \draw[very thick,black] (-3.1,-1.3) node[anchor=north west]{$\rightarrow$ from early pulse};
            \draw[very thick,black] (-3.1,-1.7) node[anchor=north west]{$\rightarrow$ from late pulse};
            \draw[very thick,black] (-3.1,-2.1) node[anchor=north west]{$\rightarrow$ from both pulses};
            \draw[very thick,black] (3,0.5) node[anchor=north west]{$\sqrt{2}t$};
            \draw[very thick,black] (0.85,-2.4) node[anchor=north west]{1};
            \draw[very thick,black] (1.35,-1.95) node[anchor=north west]{2};
            \draw[very thick,black] (1.8,-1.5) node[anchor=north west]{3};
            \draw[very thick,black] (2.3,-1) node[anchor=north west]{4};
            \draw[very thick,black] (2.75,-0.55) node[anchor=north west]{5};
            \draw[very thick,black] (0.6,-0.2) node[anchor=north west]{$t_\mathrm{A}$};
            \draw[very thick,black] (-2.3,-0.5) node[anchor=north west]{$t_\mathrm{A,-1}$};
            \draw[very thick,black] (-1.3,-0.5) node[anchor=north west]{$t_\mathrm{A,0}$};
            \draw[very thick,black] (-0.4,-0.5) node[anchor=north west]{$t_\mathrm{A,+1}$};
            \draw[very thick,black] (-3.3,3.2) node[anchor=north west]{$t_\mathrm{B}$};
            \draw[very thick,black] (-3.6,0.7) node[anchor=north west]{$t_\mathrm{B,-1}$};
            \draw[very thick,black] (-3.6,1.5) node[anchor=north west]{$t_\mathrm{B,0}$};
            \draw[very thick,black] (-3.6,2.4) node[anchor=north west]{$t_\mathrm{B,+1}$};            
    	   \end{annotate}   
         \end{subfigure}
        \caption{\textbf{Triple coincidence between the trigger and Alice's and Bob's detectors.} Histogram of the coincidence counts between pump pulse and signal and idler photon arrival times measured for an external pump power of 1mW with 200\,ps bin width and 360\,s integration time (a). The x-axis (y-axis) corresponds to photon arrival times $t_\mathrm{A}$ ($t_\mathrm{B}$) at Alice's (Bob's) detector (b). Possible detection times are $t_{-1}$, $t_{0}$ and $t_{+1}$ with respect to the pump pulse. Photons created by the two pump pulses that are 3\,ns apart appear in the five possible time-bins obtained by integrating over the antidiagonals of the histogram.}
        \label{fig:Diamond_Peaks}
\end{figure}

\noindent One sample of the measurements recorded for the time-bin tomography is shown in Fig.~\ref{fig:Diamond_Peaks}~(a).
Its interpretation is illustrated in Fig.~\ref{fig:Diamond_Peaks}~(b).
A detection time $t_{-1}$ ($t_{+1}$) represents the single photon state $\ket{1}$ ($\ket{2}$), where a photon measured at either detector took the short (long) path through both the pump and the analysis interferometers.
Photons detected at $t_0$ took the short path once and the long path once.
Integrating over the antidiagonals of the histogram gives insight into the two-photon states in five consecutive time-bins. These are illustrated as peaks in Fig.~\ref{fig:Diamond_Peaks}~(b):

\begin{center}
\begin{tabularx}{0.6\textwidth} { 
   >{\raggedright\arraybackslash}X 
   >{\centering\arraybackslash}X 
   >{\raggedleft\arraybackslash}X }

 Peak \#1 & $\ket{11}$ \\
 Peak \#2  & $\ket{1+}$, $\ket{1L}$, $\ket{+1}$, $\ket{L1}$  \\
 Peak \#3  & $\ket{++}$, $\ket{+L}$, $\ket{L+}$, $\ket{LL}$  \\
 Peak \#4  & $\ket{2+}$, $\ket{2L}$, $\ket{+2}$, $\ket{L2}$  \\
 Peak \#5  & $\ket{22}$  \\
\end{tabularx}
\end{center}

\noindent Here $\ket{+}=1/\sqrt{2}\left(\ket{1}+\ket{2}\right)$ and $\ket{L}=1/\sqrt{2}\left(\ket{1}+i\ket{2}\right)$.
The states $\ket{12}$ and $\ket{21}$ are not measured if a photon pair is created in either the early or the late pump pulse, as is the case here.

\section{Density matrix and its eigenvalues\label{app:matrix}}
The density matrix $\rho$ reconstructed via maximum likelihood estimation has positive eigenvalues demonstrating that it represents a physical state.
\\
\\
$\rho=\begin{pmatrix}
0.4961+0.j & 0.1235+0.0004j & 0.0211+0.0480j & 0.4765-0.0703j\\
0.1235-0.0004j & 0.0307+0.j      
 & 0.0053+0.0119j & 0.1185-0.0179j\\
0.0211-0.0480j & 0.0053-0.0119j & 0.0055+0.j & 0.0135-0.0491j\\
0.4765+0.0703j & 0.1185+0.0179j & 0.0135+0.0491j & 0.4676 +0.j
\end{pmatrix}$
\\
\\
Eigenvalues = (9.99999910e-01+4.48772161e-18j, 9.02487908e-08+9.08286447e-18j, 1.11913599e-10-7.45628785e-18j, 1.22826754e-13+2.73128400e-18j)
\end{appendices}

\begin{backmatter}
\bmsection{Funding}
The authors acknowledge funding by the Uniqorn project (Horizon 2020 grant agreement no. 820474), the Marie Skłodowska-Curie grant agreement No 956071 (AppQInfo) and the BeyondC project (FWF project no. F7114).

\bmsection{Author contributions}
Conceptualization, H.T., L.J., H.C., M.Kl., R.C., S.F., N.K., G.W.; Formal analysis, H.T., L.J., R.C., S.F.; Methodology,  H.T., L.J., H.C., M.Kl., R.C., S.F., N.K., G.W; Investigation, H.T., L.J., R.C., S.F.; Resources, H.S., M.Ka., S.H., C.S.; Supervision, M.Kl., R.C., S.F., C.S., N.K., G.W.; Writing - original draft, H.T., L.J.; Writing - review \& editing, All Authors; Funding acquisition, C.S., N.K., G.W.

\bmsection{Disclosures}
The authors have nothing to disclose.

\bmsection{Data availability} Data underlying the results presented in this paper are available at 10.5281/zenodo.8059483.

\end{backmatter}

\bibliography{Bibliography_TimeBin}

\end{document}